\documentclass[aps,prd,twocolumn,a4paper,superscriptaddress,10pt,floatfix]{revtex4-1}
\usepackage{graphicx}
\usepackage{multirow}
\usepackage{latexsym}
\usepackage{hyperref}
\usepackage[british]{babel}
\usepackage{amsmath,amsmath}
\usepackage{xcolor}

\begin{document}
\newcommand{\be}{\begin{equation}}
\newcommand{\ee}{\end{equation}}
\newcommand{\bq}{\begin{eqnarray}}
\newcommand{\eq}{\end{eqnarray}}

\title{Temperature-redshift relation in energy-momentum-powered gravity models}
\author{C. J. A. P. Martins}
\email{Carlos.Martins@astro.up.pt}
\affiliation{Centro de Astrof\'{\i}sica da Universidade do Porto, Rua das Estrelas, 4150-762 Porto, Portugal}
\affiliation{Instituto de Astrof\'{\i}sica e Ci\^encias do Espa\c co, Universidade do Porto, Rua das Estrelas, 4150-762 Porto, Portugal}
\author{A. M. M. Vieira}
\email{ana.mafalda.vieira@sapo.pt}
\affiliation{Faculdade de Ci\^encias da Universidade de Lisboa, Campo Grande, 1749-016 Lisboa, Portugal}

\date{\today}
\begin{abstract}
There has been recent interest in the cosmological consequences of energy-momentum-powered gravity models, in which the matter side of Einstein's equations includes a term proportional to some power, $n$, of the energy-momentum tensor, in addition to the canonical linear term. Previous works have suggested that these models can lead to a recent accelerating universe without a cosmological constant, but they can also be seen as phenomenological extensions of the standard $\Lambda$CDM, which are observationally constrained to be close to the $\Lambda$CDM limit. Here we show that these models violate the temperature-redshift relation, and are therefore further constrained by astrophysical measurements of the cosmic microwave background temperature. We provide joint constraints on these models from the combination of astrophysical and background cosmological data, showing that this power is constrained to be about $|n|<0.01$ and $|n|<0.1$, respectively in models without and with a cosmological constant, and improving previous constraints on this parameter by more than a factor of three. By breaking degeneracies between this parameter and the matter density, constraints on the latter are also improved by a factor of about two.
\end{abstract}
\maketitle
\section{\label{intro}Introduction} 

One of the phenomenological modelling routes for late-time accelerating universes is the inclusion of additional terms in the matter part of Einstein's equations. One specific such possibility is a term proportional to $T^2\equiv T_{\alpha\beta}T^{\alpha\beta}$, where $T_{\alpha\beta}$ is the energy-momentum tensor. This is known as energy-momentum-squared gravity \cite{Roshan}, but, as will be clear in what follows, it is already observationally ruled out. The model has been subsequently extended to the generic form $(T^2)^n$, dubbed energy-momentum-powered (EMP) gravity \cite{Board,Akarsu}, with the previous case corresponding to $n=1$. These models can be thought of as extensions of General Relativity with a nonlinear matter Lagrangian, and may or may not include a cosmological constant. Typically they contain two additional model parameters: the power $n$ of the nonlinear part of the Lagrangian and a normalization parameter (to be defined below) quantifying the contribution of this term to the energy budget of the universe.

The above works made the qualitative suggestion that such models can reproduce the recent acceleration of the universe without a cosmological constant. Subsequently, detailed constraints, using low-redshift background cosmology data, specifically from Type Ia supernova and Hubble parameter measurements \cite{Riess,Farooq}, were presented in \cite{Faria} and extended in \cite{Kolonia1,Kolonia2}. These works have shown that such a scenario is possible in principle (at the cost of a larger matter density than the standard $\Omega_m\sim0.3$), and also provided constraints for scenarios in which the model includes a cosmological constant, restricting them to be close to the $\Lambda$CDM limit. Broadly speaking, for the simplest scenarios in this class of models, without a cosmological constant, the power $n$ was shown to be constrained to $|n|<0.1$; if a standard cosmological constant is allowed, $n$ was not significantly constrained due to degeneracies with other model parameters (unless tight priors were imposed on these parameters).

The purpose of the present work is to show that there is a further route to constraining the low-redshift behaviour of these models. The redshift dependence of the cosmic microwave background (CMB) temperature, $T(z)=T_0(1+z)$, where $T_0= 2.7255\pm 0.0006$ K \cite{Fixsen}, is a key prediction of standard cosmology, but it is generically violated in EMP models. Current astrophysical facilities probe this temperature in the redshift range $0\le z\le6.34$, and recent work on other models \cite{Gelo,Vieira} has shown that the constraining power of these measurements is comparable to that of other background cosmology probes. Here we show that this statement also applies to EMP models. We consider scenarios with and without a cosmological constant, and show that in both cases the combination of low-redshift background cosmology and CMB temperature data significantly improves constraints on these models. Additionally, as a simple test of the robustness of these constraints, we also explore more phenomenological scenarios in which one allows for the possibility of non-standard equations of state for matter or radiation.

The plan of the rest of this work is as follows. In Sect. \ref{modelEMP} we provide a self-contained introduction to the phenomenology of EMP models, focusing on the evolution of the various components of the universe therein, with an emphasis on the radiation component, whose analysis is the main focus and novelty of this work. We then provide a brief overview of our (fully standard) analysis methods and datasets in Sect. \ref{datastat}. Sect. \ref{result} presents updated constraints on these models, with and without a cosmological constant, which are the main results of our work. As a test of their robustness, we then discuss two phenomenological extensions in Sect. \ref{pheno}. Finally, our conclusions are in Sect. \ref{conc}. We use units in which $c=1$ throughout.

\section{\label{modelEMP}The EMP models} 

In this section we start with a short review of the cosmological dynamics of EMP models, referring the reader to the original works for detailed derivations, and focusing on the evolution of the matter and radiation components, which are relevant for our subsequent analysis.

The general action for these models is assumed to take the form \cite{Roshan,Board}
\be
S=\frac{1}{2\kappa}\int\left[R+\eta (T^2)^n-2\Lambda\right]d^4x + S_{matter}\,,
\ee
where $R$ is the Ricci scalar, $\Lambda$ the cosmological constant, $\kappa=8\pi G$, and $\eta$ is a constant, whose physical units depend on the value of $n$, quantifying the overall magnitude of the phenomenological $T^2$ term. Note that a cosmological constant is in principle allowed. If this is non-zero, the model has in principle two mechanisms which may provide the recent acceleration of the universe, while if it is assumed to vanish the only available mechanism will be through the $T^2$ term.

Standard variational principles then lead to the model's Einstein equations. In a flat Friedmann-Lemaitre-Robertson-Walker universe and assuming a perfect fluid, the Friedmann, Raychaudhuri and continuity equation are
\begin{widetext}
\be
3\left(\frac{\dot a}{a}\right)^2=\Lambda+\kappa\rho+\eta(\rho^2+3p^2)^{n-1}\left[\left(n-\frac{1}{2}\right)(\rho^2+3p^2)+4np\rho\right]
\ee
\be
6\frac{\ddot a}{a}=2\Lambda-\kappa(\rho+3p)-\eta(\rho^2+3p^2)^{n-1}\left[(n+1)(\rho^2+3p^2)+4np\rho\right]
\ee
\be
{\dot\rho}=-3\frac{\dot a}{a}(\rho+p)\frac{\kappa\rho+n\eta\rho(\rho+3p)(\rho^2+3p^2)^{n-1}}{\kappa\rho+2n\eta(\rho^2+3p^2)^{n-1}\left[\left(n-\frac{1}{2}\right)(\rho^2+3p^2)+4np\rho\right]}.
\ee
\end{widetext}
As usual, the Bianchi identity implies that only two of these equations are independent. For numerical convenience in the present work, the optimal choice is to use the Friedmann and continuity equations.

\subsection{Evolution of matter}

We are interested in the low-redshift evolution of these models, in which case radiation has a negligible contribution to the Friedmann equation. We may therefore consider universes composed of ordinary matter and a cosmological constant, in which case we can simplify the Einstein equations to
\bq
3\left(\frac{\dot a}{a}\right)^2&=&\Lambda+\kappa\rho+\left(n-\frac{1}{2}\right)\eta \rho^{2n} \label{maineq1}\\
6\frac{\ddot a}{a}&=&2\Lambda-\kappa\rho-(n+1)\eta\rho^{2n}\\
{\dot\rho}&=&-3\frac{\dot a}{a}\rho\frac{\kappa+n\eta\rho^{2n-1}}{\kappa+(2n-1)n\eta\rho^{2n-1}}. \label{maineq2} 
\eq
We note that the Friedmann equation has some phenomenological similarities with the Cardassian model \cite{Cardassian}. A simple inspection of the equations leads to the expectation that $n>1/2$ might have nontrivial impacts at early times, but will become irrelevant at more recent times. Conversely, $n<1/2$ may be interesting at late times, in particular as a possible explanation for the recent acceleration of the universe. In passing, we also note that there are three specific choices of $n$ for which these equation can be solved analytically in a low-redshift limit, all of which have been mathematically studied: $n=1$ in \cite{Roshan}, $n=1/2$ in \cite{Early}, and $n=0$ in \cite{Board} respectively. 

In what follows we do not restrict ourselves to specific choices of $n$, but consider the general case in which $n$ is a free parameter, and these equations, therefore, need to be solved numerically. It is convenient to define a dimensionless cosmological density $r(z)$, via $\rho (z)= \rho_0 r(z)$, where $\rho_0$ is the present day density, as well as a generic parameter
\be
Q=\frac{\eta}{\kappa}\rho_0^{2n-1}\,.
\ee
With these assumptions the continuity equation for a model containing matter and a cosmological constant, previously introduced in Eq. (\ref{maineq2}), and expressed in terms of redshift instead of time, has the form
\be
\frac{dr}{dz}=\frac{3r}{1+z}\times\frac{1+nQr^{2n-1}}{1+(2n-1)nQr^{2n-1}}\,;
\ee
this can be numerically integrated to yield $r(z)$, and the result can then be substituted into the Friedmann equation, previously defined in Eq. (\ref{maineq1}), which becomes
\be
E^2(z)=\frac{H^2(z)}{H_0^2}=\Omega_\Lambda+\Omega_mr+\left(n-\frac{1}{2}\right)Q\Omega_mr^{2n}\,.
\ee
Here we have used the standard definitions
\bq
\Omega_\Lambda &=& \frac{\Lambda}{3H_0^2}\\
\Omega_m &=& \frac{\kappa\rho_0}{3H_0^2}\,.
\eq
Our flatness assumption requires that
\be
\Omega_\Lambda=1-\left[1+\left(n-\frac{1}{2}\right)Q\right]\Omega_m\,,
\ee
and therefore the Friedmann equation can also be written in the two alternative forms
\bq
E^2(z)&=&\Omega_\Lambda+\Omega_mr+(1-\Omega_m-\Omega_\Lambda)r^{2n}\\
E^2(z)&=&1+\Omega_m(r-1)+\left(n-\frac{1}{2}\right)Q\Omega_m(r^{2n}-1)\,;
\eq
the first one is generic, while the second applies only if $\Omega_\Lambda \neq0$. On the other hand, if $\Omega_\Lambda=0$ we can also use the flatness assumption to eliminate $Q$ in the continuity equation, writing it as
\be
\frac{dr}{dz}=\frac{3r}{1+z}\times\frac{(2n-1)\Omega_m+2n(1-\Omega_m)r^{2n-1}}{(2n-1)[\Omega_m+2n(1-\Omega_m)r^{2n-1}]}\,.
\ee

One sees that in these models the recent acceleration of the universe could stem from the nonlinear term in a matter-only universe with $n=0$, while having the canonical vacuum energy density $\Omega_\Lambda=0$. We will return to this point in the conclusions. For $n$ close to but not equal to zero, there are two effects. Firstly, the (formerly) constant term in the Friedmann equation becomes slowly varying, and secondly, the continuity equation implies that the matter density does not behave exactly as $r\propto (1+z)^3$. Both of these have observational consequences, and therefore lead to constraints on $n$.

\subsection{Evolution of radiation and other fluids}

More generally, we can  consider fluids with a constant equation of state $w=p/\rho=$const. In this case the continuity equation becomes
\be
\frac{dr}{dz}=\frac{3r}{1+z}(1+w)\times\frac{1+nQf_1(n,w)r^{2n-1}}{1+2nQf_2(n,w)r^{2n-1}}\,,
\ee
where for convenience we have defined the dimensionless functions
\bq
f_1(n,w) &=& (1+3w)(1+3w^2)^{n-1}\,, \label{deff1}\\
f_2(n,w) &=& (1+3w^2)^{n-1}\left[\left(n-\frac{1}{2}\right)(1+3w^2)+4nw\right]\,. \label{deff2}
\eq
Now the Friedmann equation has the form
\be
E^2(z)=\Omega_\Lambda+\Omega_mr+f_2(n,w)Q\Omega_mr^{2n}\,,
\ee
together with the consistency relation
\be
\Omega_\Lambda=1-\left(1+f_2Q\right)\Omega_m\,.
\ee
It is worthy of note that this has the same explicit generic form as before (although of course the redshift dependence is different),
\bq
E^2(z)&=&\Omega_\Lambda+\Omega_mr+(1-\Omega_m-\Omega_\Lambda)r^{2n}\\
E^2(z)&=&1+\Omega_m(r-1)+f_2(n,w)Q\Omega_m(r^{2n}-1)\,,
\eq
where, again, the first is generic while the second applies for $\Omega_\Lambda\neq0$. On the other hand, if $\Omega_\Lambda=0$ the continuity equation can also be written in a way that eliminates the parameter $Q$,
\be
\frac{dr}{dz}=\frac{3r}{1+z}(1+w)\times\frac{\Omega_mf_2+n(1-\Omega_m)f_1r^{2n-1}}{f_2[\Omega_m+2n(1-\Omega_m)r^{2n-1}]}\,.
\ee

Of specific interest is the case of radiation. that is $w=1/3$, for which we can write
\be
\frac{dr}{dz}=\frac{4r}{1+z}\times\frac{1+2(4/3)^{n-1}nQr^{2n-1}}{1+2(4/3)^n(2n-1/2)nQr^{2n-1}}\,,
\ee
with a corresponding temperature-redshift relation
\be
T(z)=T_0\, r(z)^{1/4}\,.
\ee
This differs from the standard temperature-redshift relation, $T(z)=T_0(1+z)$, except in two specific cases: the trivial $n=0$ and the rather less trivial $n=5/8$. The physical assumptions underlying the standard relation are that the CMB spectrum was originally a black-body, the expansion of the Universe is adiabatic, and the photon number is conserved. Thus the EMP models are, phenomenologically, similar to others where additional physical mechanisms imply deviations from the canonical behaviour \cite{Avgoustidis,Euclid}. One may even phenomenologically generalize this relation for the case where the radiation equation of state, $w_r$ is close to but not exactly $1/3$, which which case we have
\be\label{genrel}
T(z)=T_0\, r(z)^{w_r/(1+w_r)}\,.
\ee
We note that for such changes to the temperature-redshift relation to be observationally plausible the putative photon production/destruction processes are required to be adiabatic and achromatic (i.e., not dependent on frequency). This 'benign' case is the only one worth considering: outside it one would have CMB spectral distortions which would rule out the model \cite{Chluba}.

Incidentally, we note that there is a third value of $n$ for which the continuity equation has an analytic but non-standard solution: for $n=1/2$, we have
\be
T(z)=T_0(1+z)^\lambda\,,\qquad \lambda=\frac{1+\sqrt{3}Q/2}{1+Q/\sqrt{3}}\,,
\ee


\section{\label{datastat}Available data and analysis method}

Our goal is to update the constraints on EMP models, previously discussed in \cite{Faria,Kolonia1,Kolonia2} and based on background cosmology data. Specifically we will  combine this cosmological data with astrophysical measurements of the CMB temperature.

The constraints in the earlier works relied on two background cosmology datasets, which we use again here for the dual purpose of providing a fair comparison and of validating our analysis codes. The first of these datasets is the Pantheon compilation \cite{Scolnic,Riess}. This is a data set of 1048 Type Ia supernovae, compressed into 6 correlated measurements of $E^{-1}(z)$ and spanning the redshift range $0.07<z<1.5$. (Strictly speaking, this compression also relies on 15 Type Ia supernovae from two Hubble Space Telescope Multi-Cycle Treasury programs, and it assumes a spatially flat universe.) The compression methodology and validation are detailed in Section 3 of  \cite{Riess}. The second dataset is the compilation of 38 Hubble parameter measurements of Farooq {\it et al.} \cite{Farooq}: this is a more heterogeneous set includes both data from cosmic chronometers and from baryon acoustic oscillations. Both of these are canonical data sets and have been extensively used in the literature. Together, the two cosmological data sets contain measurements up to redshift $z\sim2.36$, and when using the two in combination we will refer to this as the cosmological data.

Measurements of the CMB temperature at nonzero redshift can be obtained from two different observational techniques. A total of 45 such measurements have been compiled in \cite{Gelo}, and will be used here. At low redshifts (typically $z<1$), one can use the thermal Sunyaev-Zel'dovich (SZ) effect. Currently available measurements come from 815 Planck clusters in 18 redshift bins \cite{Hurier} and 158 SPT clusters in 12 redshift bins \cite{Saro}. In the approximate range $1<z<3$ one can rely on high-resolution spectroscopy of molecular or atomic species whose energy levels can be excited by CMB photons. The first such measurement (as opposed to an upper limit) was obtained in the year 2000 by Srianand {\it et al.} \cite{Srianand}, and the recent work of \cite{Klimenko} updates various earlier measurements of this type. The redshift range of the CMB temperature measurements largely overlaps with that of the cosmological data. Strictly speaking the redshift range of the former has recently been enlarged to reach $z\sim6.34$ \cite{Riechers}; this is a relatively weak constraint (therefore with very little statistical weight), but we do include it for completeness.

We use a standard statistical likelihood analysis, as described e.g. in \cite{Verde}. The likelihood is defined as
\be
{\cal L}(p)\propto\exp{\left(-\frac{1}{2}\chi^2(p)\right)}\,,
\ee
where $p$ symbolically denotes the free parameters in the model under consideration. The chi-square for a relevant redshift-dependent quantity $O(z)$ has the explicit form
\be
\chi^2(p)=\sum_{i,j}\left(O_{obs,i}-O_{mod,i}(p)\right)C_{ij}^{-1}\left(O_{obs,j}-O_{mod,j}(p)\right)\,,
\ee
where the obs and mod subscripts denote observations and model respectively, and $C$ is the covariance matrix of the dataset (which may, in some cases, be trivial).

In all our analyses the Hubble constant is never used as a free parameter; instead it is always analytically marginalized, following the procedure described in \cite{Homarg}. Since one can trivially write $H(z)=H_0 E(z)$, one notices that $H_0$ is purely a multiplicative constant, and can be analytically integrated in the likelihood. Towards this end one computes quantities
\bq
A(p)&=&\sum_{i}\frac{E_{mod,i}^2(p)}{\sigma^2_i}\\
B(p)&=&\sum_{i}\frac{E_{mod,i}(p) H_{obs,i}}{\sigma^2_i}\\
C(p)&=&\sum_{i}\frac{H_{obs,i}^2}{\sigma^2_i}
\eq
where the $\sigma_i$ are the uncertainties in observed values of the Hubble parameter. Then, the chi-square is given by
\be
\chi^2(q)=C(q)-\frac{B^2(q)}{A(q)}+\ln{A(q)}-2\ln{\left[1+Erf{\left(\frac{B(q)}{\sqrt{2A(q)}}\right)}\right]}
\ee
where $Erf$ is the Gauss error function and $\ln$ is the natural logarithm. It follows that our results do not depend on possible choices of the Hubble constant, and are therefore unaffected by to the so-called Hubble tension. 

Moreover, the present-day value of the CMB temperature, $T_0$, also appears purely as a multiplicative factor in the temperature-redshift relation. The same remarks of the previous paragraph therefore also hold for it, and is also analytically marginalized in what follows,

Our analyses are grid-based, using Matlab and Python codes that have been independently custom-built for this work, but validated against each other and also by comparison with results in previous works. Unless otherwise is stated, we use uniform priors for the model parameters, in the plotted ranges. We have tested that these assumptions do not significantly impact our results.

\section{\label{result}Joint low-redshift constraints}

We now proceed to present our main results, specifically the constraints on these models, with or without a cosmological constant allowed. Two further extensions of the case without a cosmological constant are also considered in the next section. In all cases we report both the separate constraints from the cosmology and the CMB temperature data sets, introduced in the previous section, and the joint constraints from the combined data set. Table \ref{table1} summarizes these results.

\begin{table*}
\centering
\caption{One sigma ($\Delta\chi^2=1$) constraints on the power $n$, the matter density $\Omega_m$ and a third free parameter (when applicable) for various flat EMP models. The specific assumptions for each set of constraints are described in the main text. We report separate constraints from the cosmological and CMB data described in Sect. \ref{datastat}, as well as the joint constraints from the combination of the two.}
\begin{tabular}{| c | c | c | c | c |}
\hline
Model assumptions & Data & $\Omega_m$ & $n$  & Other \\
\hline
{} & Cosmo & $0.39^{+0.06}_{-0.08}$ & $0.04^{+0.03}_{-0.04}$ & N/A \\
$\Omega_\Lambda=0$ (Sect. \ref{model1}) & $T_{CMB}$ & Unconstrained & $0.004^{+0.006}_{-0.004}$ & N/A \\
{} & Joint & $0.30\pm0.02$ & $0.005\pm0.005$ & N/A \\
\hline
{} & Cosmo & $0.29\pm0.03$ & Unconstrained & Unconstrained  \\
$\Omega_\Lambda\neq0$ (Sect. \ref{model2})  & $T_{CMB}$ & Unconstrained & $[-0.06,+0.04]$ & $\Omega_\Lambda=0.76^{+0.14}_{-0.22}$ \\
{} & Joint & $0.29\pm0.02$ &  $[-0.07,+0.05]$ & $\Omega_\Lambda=0.77^{+0.08}_{-0.17}$ \\
\hline
{} & Cosmo & $0.30^{+0.11}_{-0.10}$ & $-0.07^{+0.06}_{-0.04}$ & $w_m=-0.11^{+0.06}_{-0.03}$ \\
$w_m=const.$ (Sect. \ref{model3})  & $T_{CMB}$ & Unconstrained & $0.005^{+0.004}_{-0.005}$ & Unconstrained \\
{} & Joint & $0.34^{+0.05}_{-0.04}$ & $0.006\pm0.006$ & $w_m=-0.06^{+0.04}_{-0.03}$ \\
\hline
{} & Cosmo & $0.39^{+0.06}_{-0.08}$ & $0.04^{+0.03}_{-0.04}$ & Unconstrained \\
$w_r=const.$ (Sect. \ref{model4})  & $T_{CMB}$ & Unconstrained & $0.01\pm0.01$ & $w_r=0.347\pm0.015$ \\
{} & Joint & $0.33\pm0.03$ & $0.02\pm0.01$ & $w_r=0.352\pm0.015$ \\
\hline
\end{tabular}
\label{table1}
\end{table*}

\subsection{\label{model1}Ordinary fluids without a cosmological constant}

The simplest case is the one where the model is taken as a genuine alternative to the canonical $\Lambda$CDM model, by setting $\Omega_\Lambda=0$ and further assuming that the matter and radiation fluids have the standard equations of state, $w_m=0$ and $w_r=1/3$ respectively. In this case the parameter space in our analysis has two free parameters, $(\Omega_m,n)$, and it suffices to consider top-hat priors in the ranges $\Omega_m\in[0.15,0.45]$ and $n\in[-0.1,0.1]$.

\begin{figure*}
  \begin{center}
    \includegraphics[width=\columnwidth]{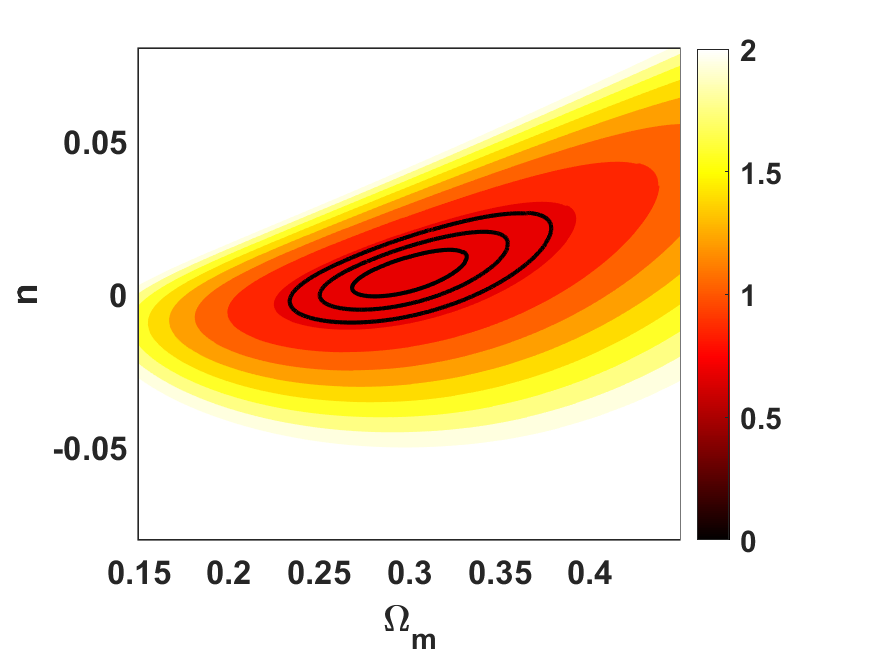}
    \includegraphics[width=\columnwidth]{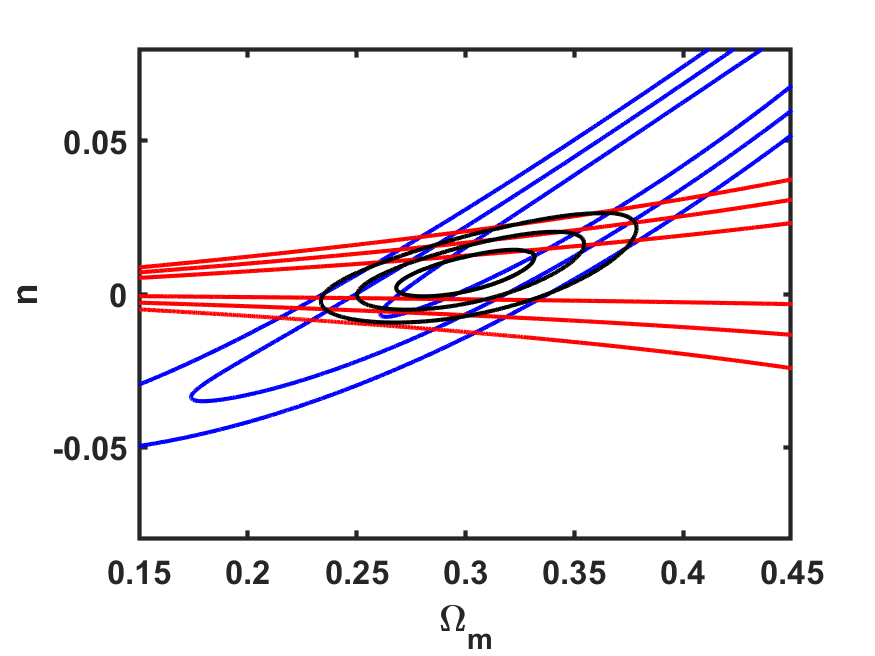}
    \includegraphics[width=\columnwidth]{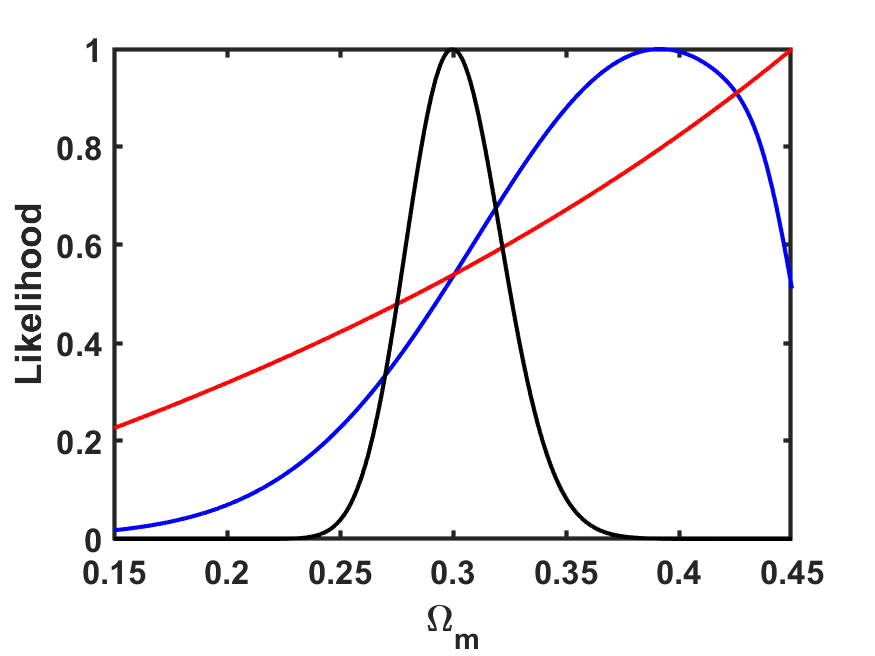}
    \includegraphics[width=\columnwidth]{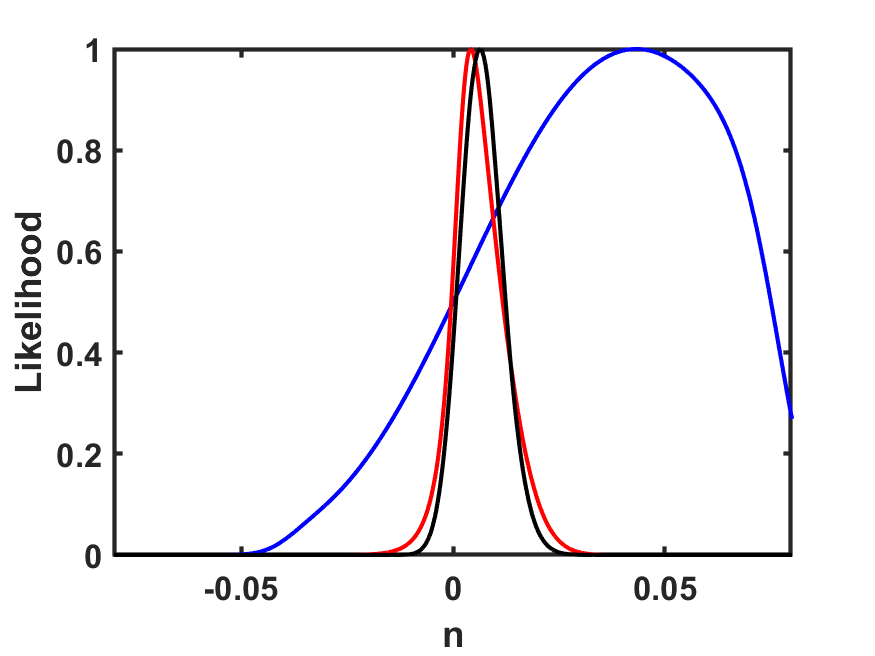}
    \caption{Constraints on the EMP model for flat universes with standard fluids and $\Omega_\Lambda=0$; the top and bottom panels show the 2D and 1D (marginalized) constraints respectively. Blue, red and black lines correspond to one, two and three sigma constraints from cosmology, CMB and the joint data sets respectively, and the colormap depicts the reduced chi.square.}
    \label{fig01}
  \end{center}
\end{figure*}

Figure \ref{fig01} and the first set of rows of Table \ref{table1} show the results of this analysis, and clearly illustrate the synergies between the two data sets. As previously reported \cite{Faria}, for the cosmological data there is a degeneracy between the two parameters, leading to relatively weak constraints on each of them. The best-fit values of both differ by about one standard deviation from the canonical values, and in particular a slightly higher matter density would be preferred.

The addition of the CMB temperature measurements breaks this degeneracy. They have very little sensitivity to the matter density but are very sensitive to $n$, tightening its constraint by almost an order of magnitude. When combining the two datasets we essentially recover the canonical $\Lambda$CDM behaviour, with the standard preferred value for the matter density and the power of the nonlinear part of the Lagrangian constrained to be approximately $|n|<0.01$.

\subsection{\label{model2}Ordinary fluids with a cosmological constant}

Here we keep the standard behaviour for the matter and radiation fluids, but allow for the possibility of a non-zero cosmological constant, $\Omega_\Lambda\neq0$. Clearly in this case there are two terms in the Einstein equation which can provide a low-redshift accelerating universe, and therefore one expects further degeneracies in the model parameters. We confirm that this is so, but nevertheless the combination of the cosmology and CMB temperature data can still provide meaningful constraints on the model.

\begin{figure*}
  \begin{center}
    \includegraphics[width=0.68\columnwidth]{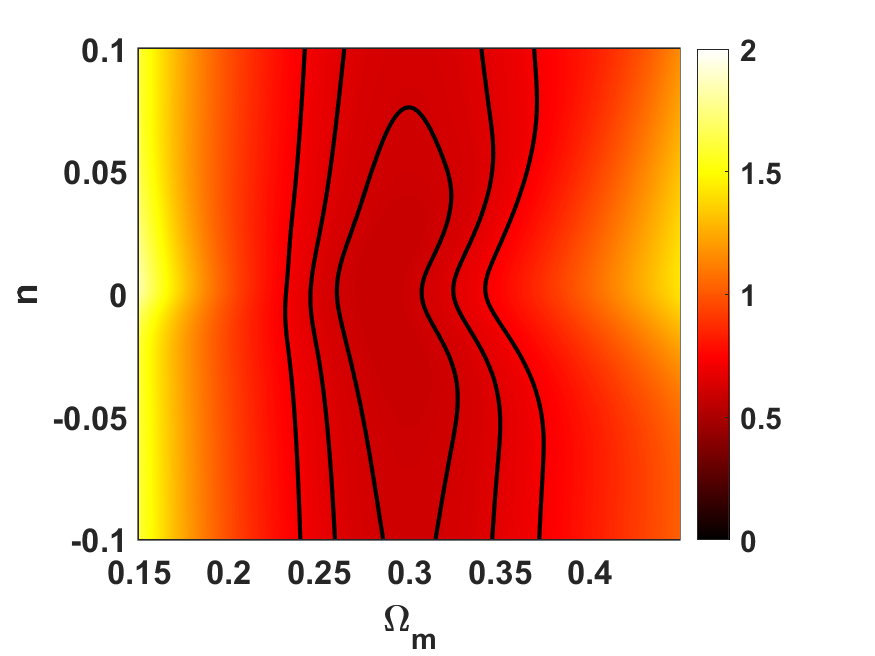}
    \includegraphics[width=0.68\columnwidth]{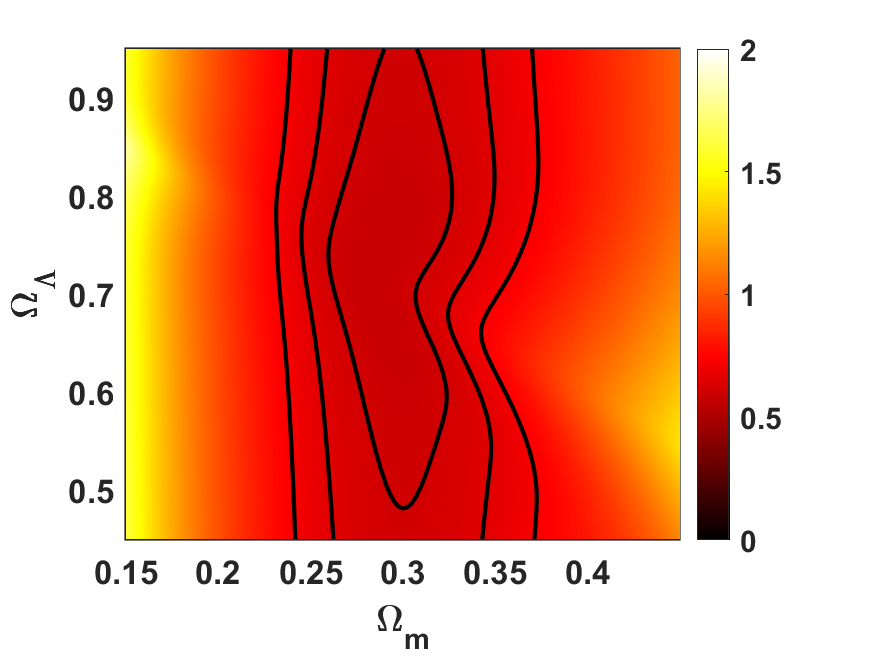}
    \includegraphics[width=0.68\columnwidth]{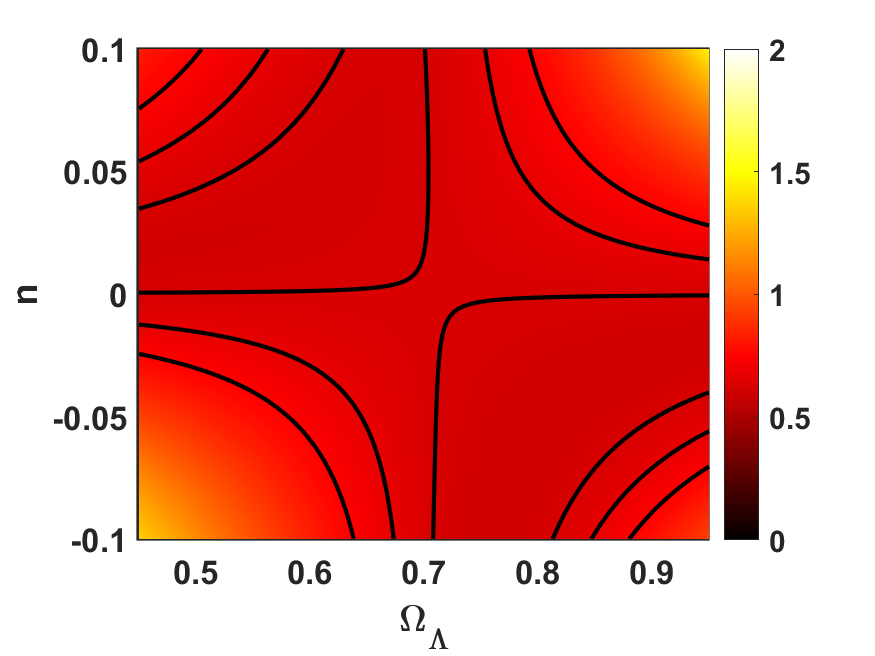}
    \includegraphics[width=0.68\columnwidth]{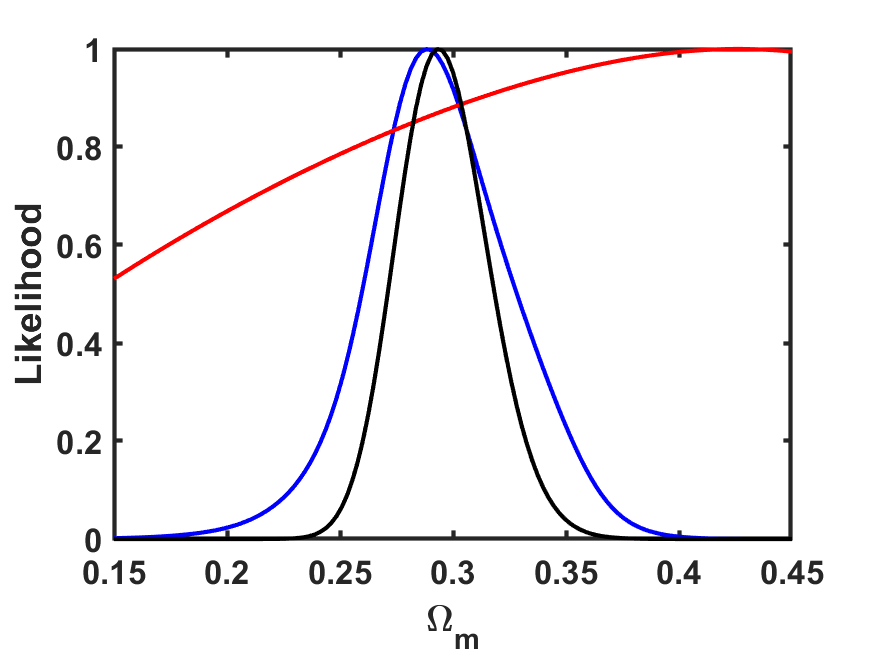}
    \includegraphics[width=0.68\columnwidth]{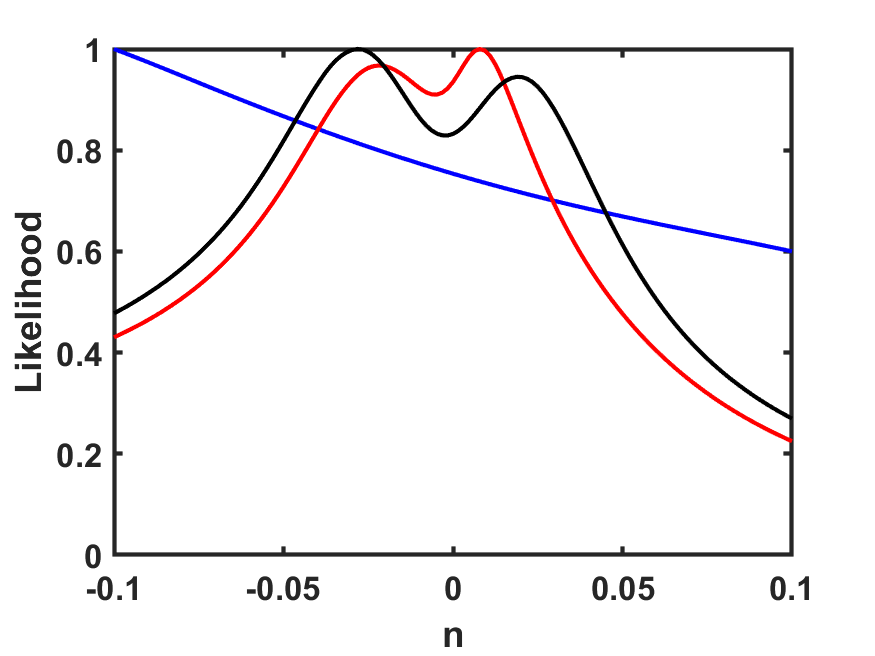}
    \includegraphics[width=0.68\columnwidth]{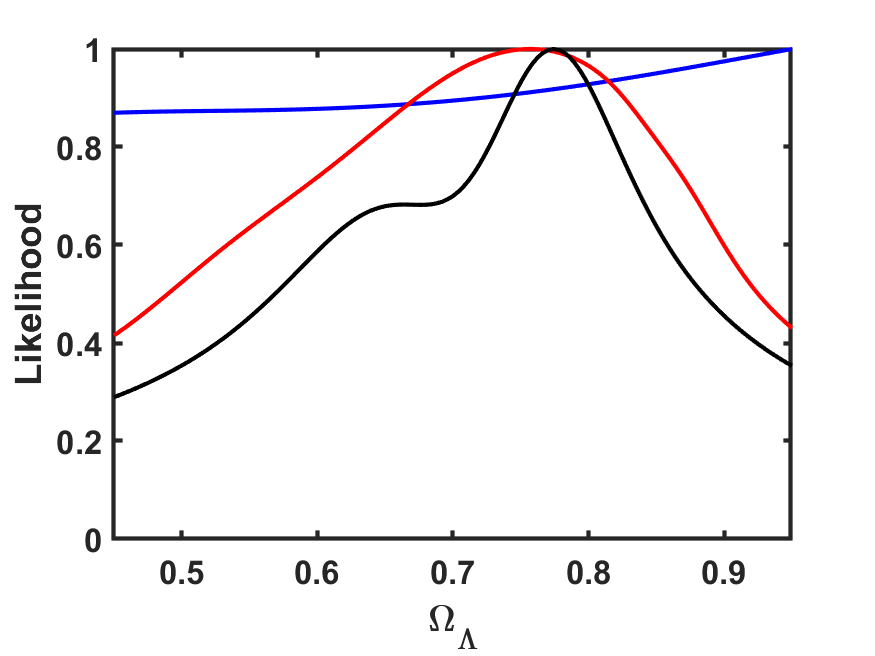}
    \caption{Constraints on the EMP model for flat universes with standard fluids and $\Omega_\Lambda\neq0$; the top and bottom panels show the 2D and 1D (marginalized) constraints respectively. Blue, red and black lines correspond to one, two and three sigma constraints from cosmology, CMB and the joint data sets respectively, and the colormap depicts the reduced chi.square.}
    \label{fig02}
  \end{center}
\end{figure*}

In this case, our parameter space has three free parameters, and we have a choice between two physically meaningful options: $(\Omega_m,n,Q)$ and $(\Omega_m,n,\Omega_\Lambda)$, In what follows we use the latter, but we have checked that the alternative choice would yield comparable constraints for the common parameters, $n$ and $\Omega_m$---specifically, the differences would be of the other of ten percent or less. We keep the priors for the matter density and the power of the nonlinear term stated in the previous subsection, while for the cosmological constant we take the top-hat prior $\Omega_\Lambda\in[0.45,0.95]$.

Figure \ref{fig02} and the second set of rows of Table \ref{table1} show the results of this analysis, The degeneracy between the two available acceleration mechanisms, specifically between $\Omega_\Lambda$ and $n$, implies that the cosmological data can only constrain the matter density, which unsurprisingly is consistent with the canonical value. As in the previous case, the CMB temperature partially breaks this degeneracy and yields meaningful constraints on the two parameters, which are further improved in the joint analysis (which also tightens the constraint on the matter density). We also note that the posterior likelihood for $n$ is highly non-Gaussian, so for this case we report the range of values within a $\Delta\chi^2=1$ of the best-fit value.

In this case the constraint on $n$ is weakened by about a factor of ten with respect to the case without a cosmological constant, but it is still a tight $|n|<0.1$, comfortably ruling out the phenomenologically simple case $n=1/2$, which naively one might have expected to be the most natural value in this class of models. On the other hand, the preferred value of $\Omega_\Lambda$ is statistically consistent with the standard one, implying that the contribution of the nonlinear term is subdominant with respect to the standard one.

\section{\label{pheno}Further phenomenological extensions}

The previous section has shown that this class of model is tightly constrained to the neighborhood of $\Lambda$CDM limit, with the parameter $n$ very close to zero. In particular, the original energy-momentum-squared gravity (corresponding to $n=1$, is evidently ruled out). In this section, we briefly explore how the previous constraints are affected by enlarging the parameter space, focusing on the $\Lambda=0$ case. 

Specifically, we consider extended models, where either matter or radiation have an equation of state which slightly deviates from their standard values. It is clear that these scenarios are more phenomenological than those discussed in the previous section, and that such deviations in the equation of state are ab initio constrained to be quite small. Nevertheless, these analyses serve the purpose of providing simple robustness tests of our main constraints, in a sense further described in what follows.

\subsection{\label{model3}Non-standard matter content}

We now return to the case without a cosmological constant, $\Omega_\Lambda=0$, but relax the assumption of a standard matter component. Specifically, instead of assuming a matter equation of state $w_m=0$ we promote it to a third free parameter, while requiring it to have a constant value. It should be remarked that the equation of state of matter is tightly constrained by several previous works, e.g. \cite{Tutusaus,Ilic} report constraints at the level of $|w_m|<0.003$. Our goal here is to see how analogous constraints in this case compare to the previously reported ones.

\begin{figure*}
  \begin{center}
    \includegraphics[width=0.68\columnwidth]{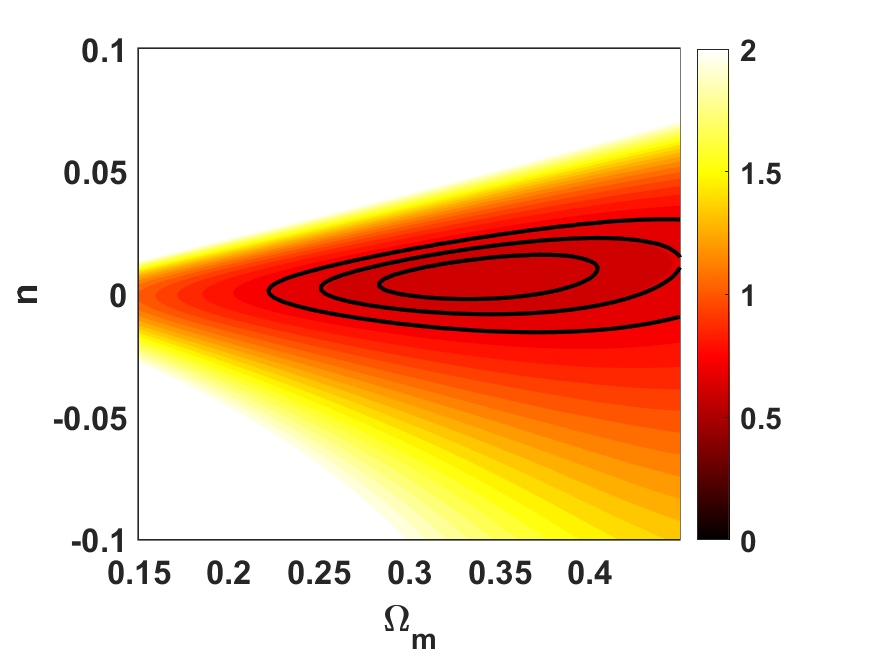}
    \includegraphics[width=0.68\columnwidth]{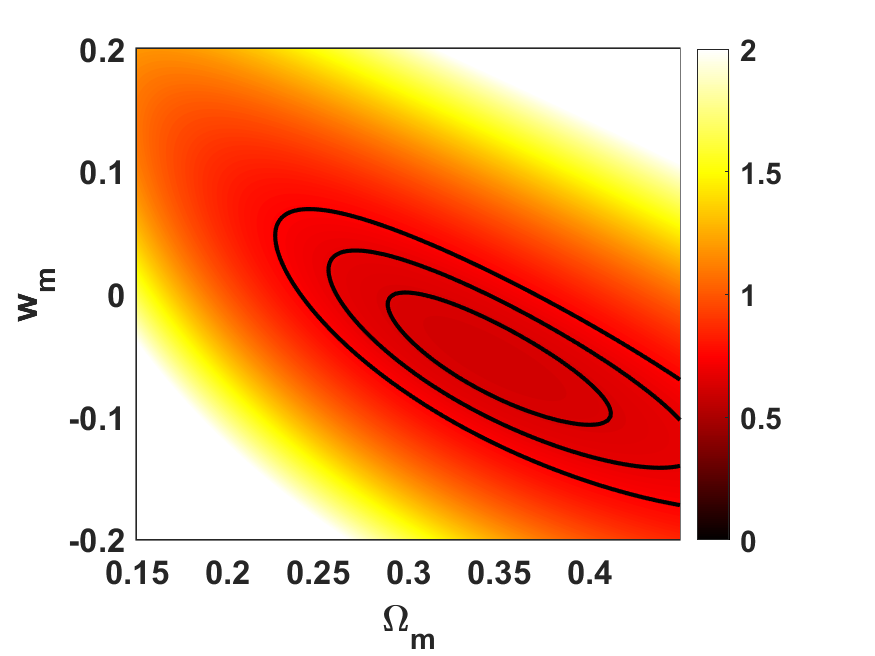}
    \includegraphics[width=0.68\columnwidth]{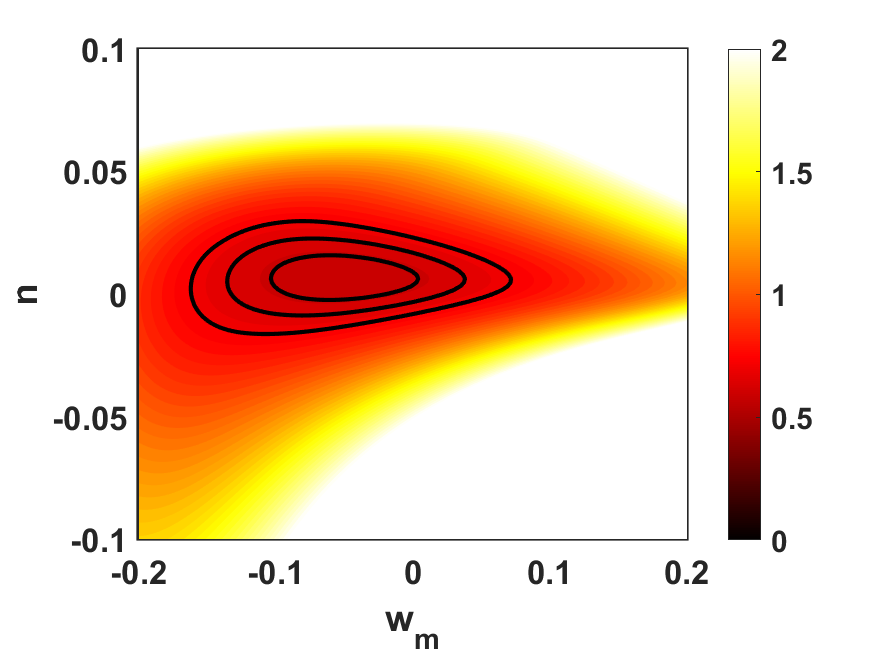}
    \includegraphics[width=0.68\columnwidth]{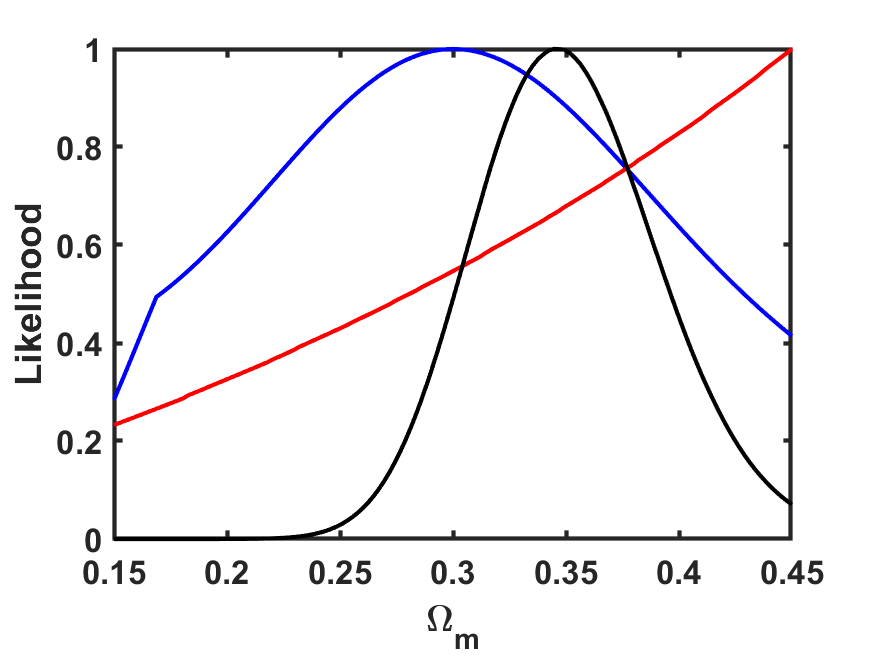}
    \includegraphics[width=0.68\columnwidth]{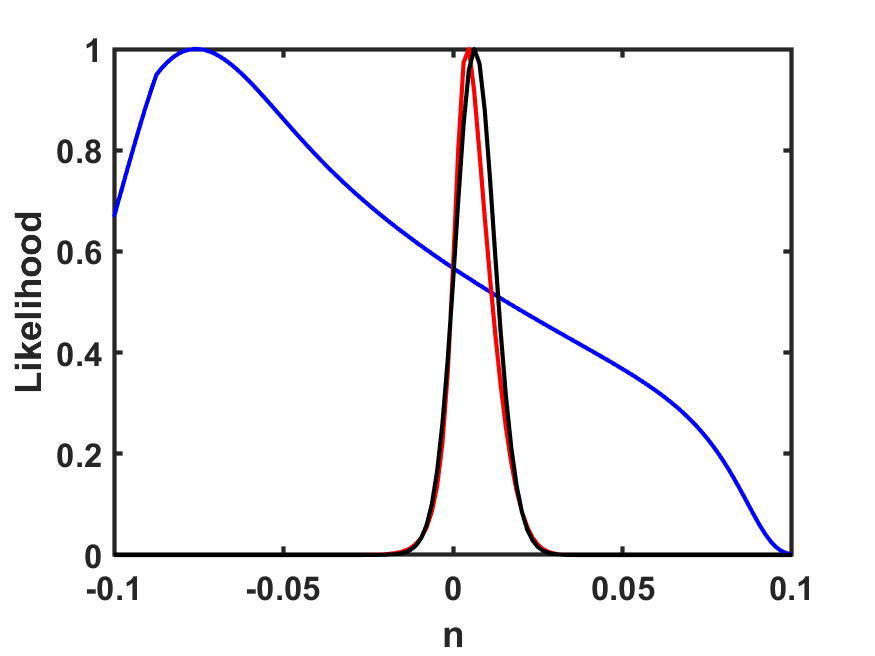}
    \includegraphics[width=0.68\columnwidth]{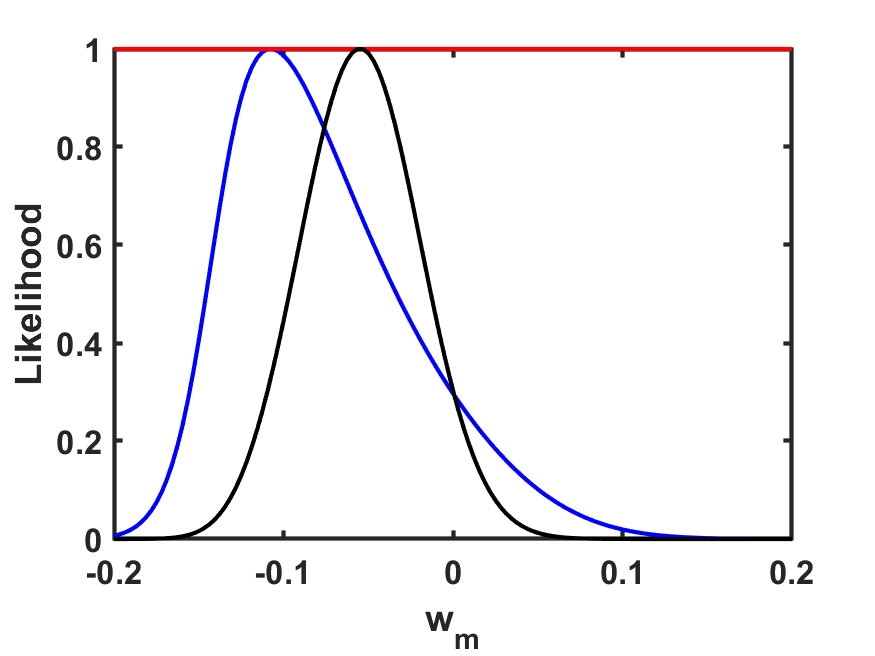}
    \caption{Constraints on the EMP model for flat universes with $\Omega_\Lambda=0$ and $w_m=const$; the top and bottom panels show the 2D and 1D (marginalized) constraints respectively. Blue, red and black lines correspond to one, two and three sigma constraints from cosmology, CMB and the joint data sets respectively, and the colormap depicts the reduced chi.square.}
    \label{fig03}
  \end{center}
\end{figure*}

In this case we also have a three-dimensional parameter space to constrain, specifically $(\Omega_m,n,w_m)$. We again keep the previously mentioned priors for the first two parameters, while for the matter equation of state we assume a top-hat prior with a range $w_m\in[-0.2,0.2]$. In other words, this is an extension of the case reported in Sect. \ref{model1}.

Figure \ref{fig03} and the third set of rows of Table \ref{table1} show the results of this analysis, Taking first the cosmological data, the constraint on the matter density becomes quite weak, though interestingly the best-fit value is now the standard one---which wasn't the case in Sect. \ref{model1}. This shift in the best-fit value is compensated by a small preference for negative values for $n$ and $w_m$. In both cases this preference is at a level smaller than two standard deviations, and therefore it is not statistically significant. As in the previous case, the CMB temperature data has very little sensitivity to the matter density (and obviously none at all to its equation of state), but it constrains $n$ ten times more strongly than the cosmology data, again breaking some of the parameter degeneracies.

The end result, in the joint constraints, is fairly similar to the one obtained above for  the $w_m=0$ case, The constraint on $n$ is almost unchanged, while that for the matter density becomes weaker, and the best-fit value also shifts slightly, by about one standard deviation (which is not statistically significant). There is still a less than two sigma preference (again, not statistically significant) for a negative value of $w_m$, while for $n$ the posterior distribution is now skewed towards slightly positive values. Finally, our uncertainties on the matter equation of state are about a factor of ten weaker than those obtained in \cite{Tutusaus,Ilic} for more standard models, though it should also be noticed that these works also use higher redshift data, specifically from CMB temperature and polarization data.

This analysis therefore leads to infer that even if a matter component with an 'extreme' equation of state $w_m\sim-0.1$ were allowed, $n$ would still be consistent with zero, and the component responsible for accelerating the universe would still effectively behave as a cosmological constant. It's in this sense that this analysis is a robustness test of the earlier constraints: one can't reasonably concoct a preferred value of $n$ differing from the $\Lambda$CDM limit by changing the matter constant.

\subsection{\label{model4}Non-standard radiation content}

Finally, we take one more case, analogous to the one in the previous subsection. In this case we assume the standard behaviour for matter and no cosmological constant, but allow for a constant equation of state for radiation, not necessarily with the standard value of $w_r=1/3$. Such scenarios have also been explored in other models where photon number is not conserved \cite{Gelo,Jetzer1,Jetzer2}. Phenomenologically, and for values of $w_r$ close to the standard one, the generalized temperature-redshift relation is given by Eq. (\ref{genrel}), keeping in mind that this further assumes that the underlying physical photon production/destruction processes (which we leave unspecified) are adiabatic and achromatic: if this were not the case one would have CMB spectral distortions which would rule of the model.

With these caveats we again have a three-dimensional parameter space to constrain, $(\Omega_m,n,w_r)$, and we keep the previously mentioned priors for the first two parameters, while for the radiation equation of state we assume a top-hat prior with a range $w_r\in[0.3,0.4]$. This is also an extension of the case reported in Sect. \ref{model1}.

\begin{figure*}
  \begin{center}
    \includegraphics[width=0.68\columnwidth]{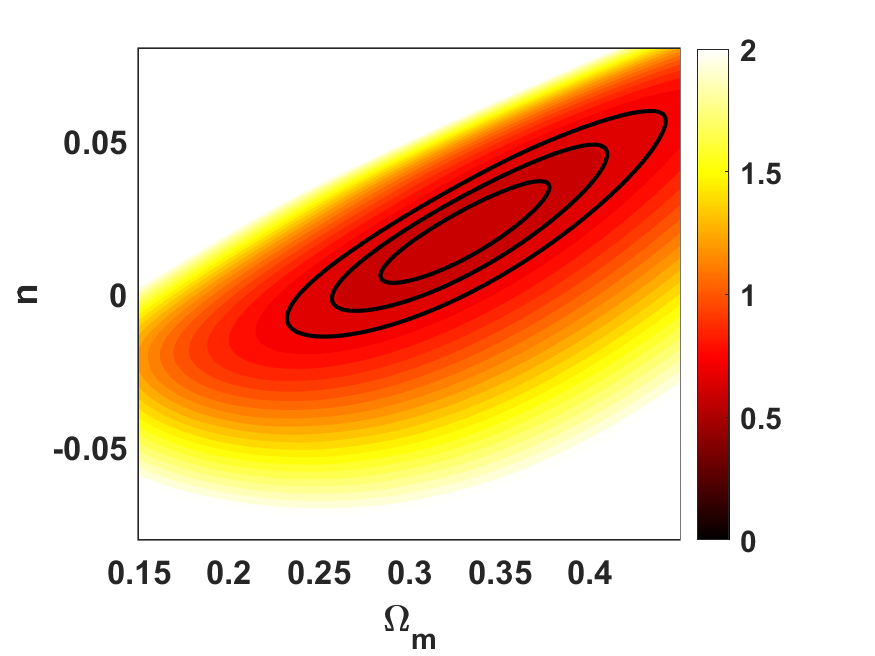}
    \includegraphics[width=0.68\columnwidth]{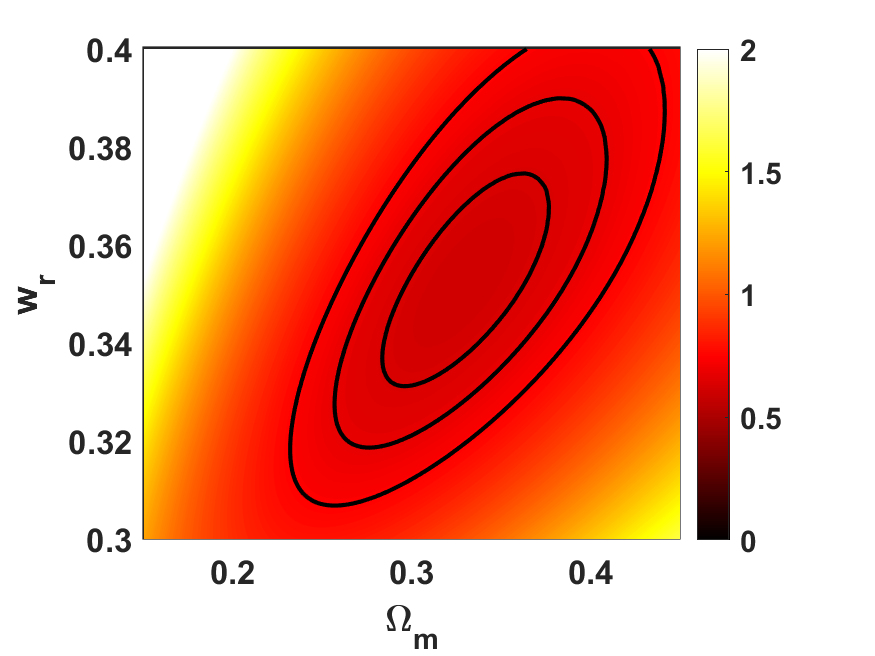}
    \includegraphics[width=0.68\columnwidth]{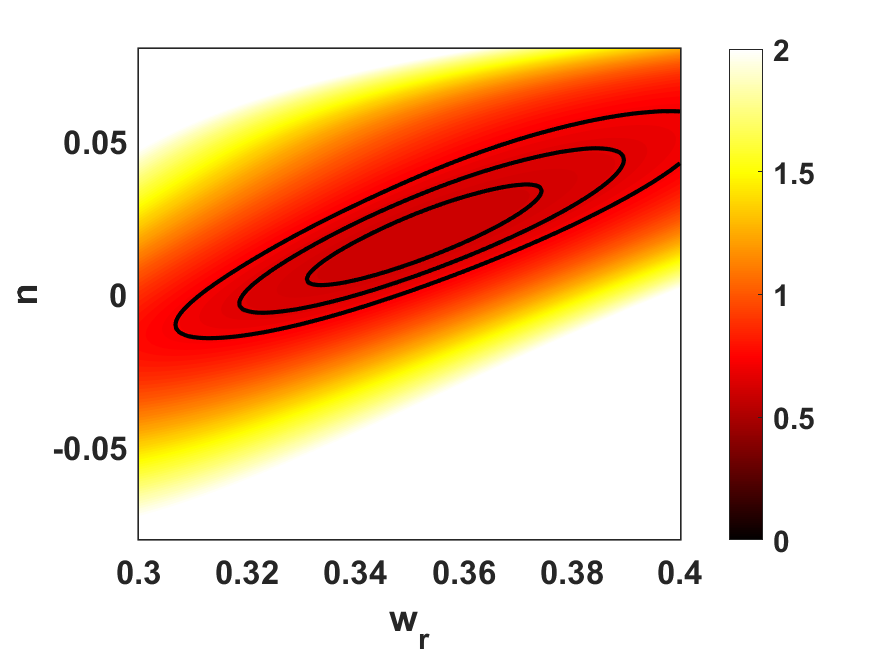}
    \includegraphics[width=0.68\columnwidth]{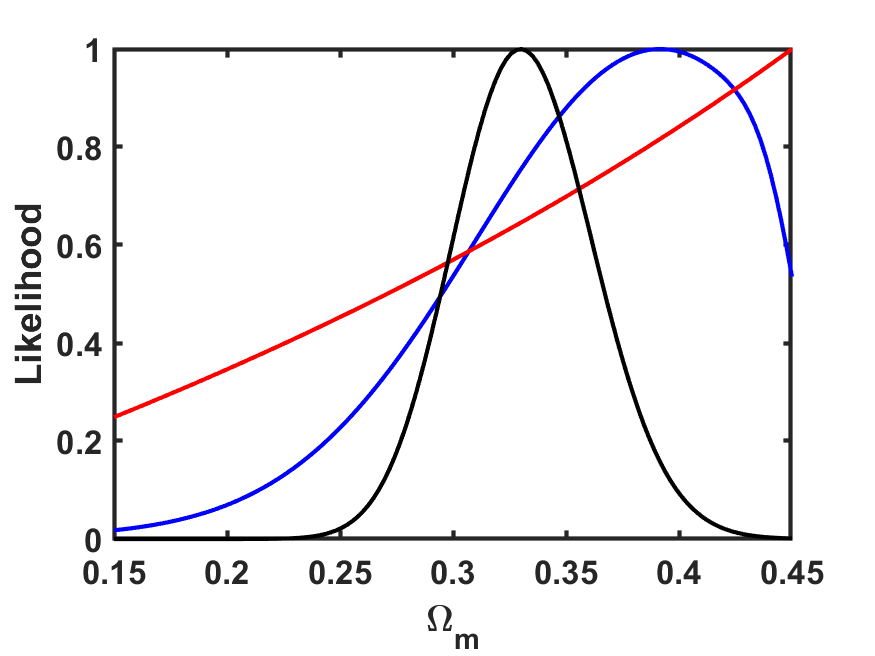}
    \includegraphics[width=0.68\columnwidth]{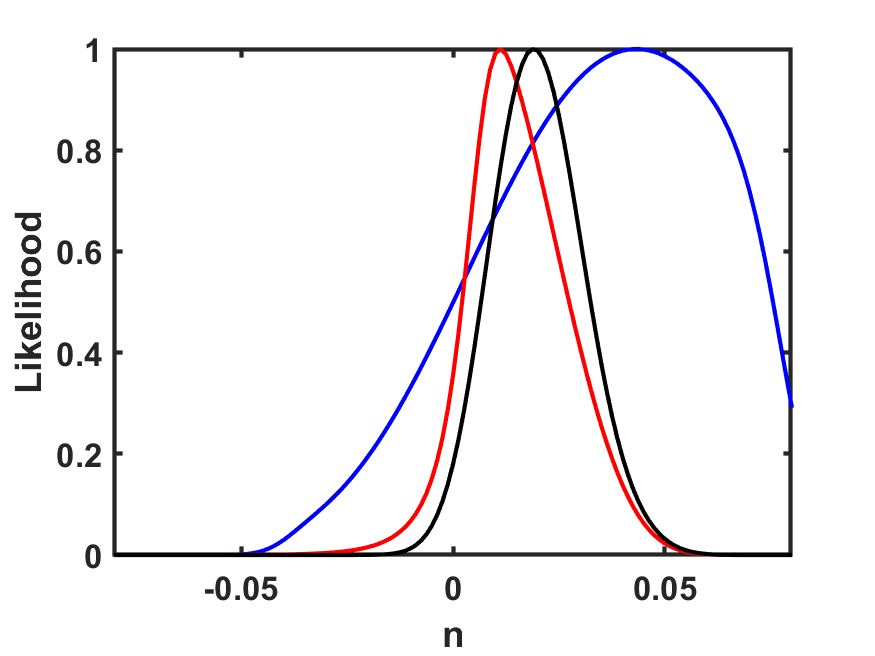}
    \includegraphics[width=0.68\columnwidth]{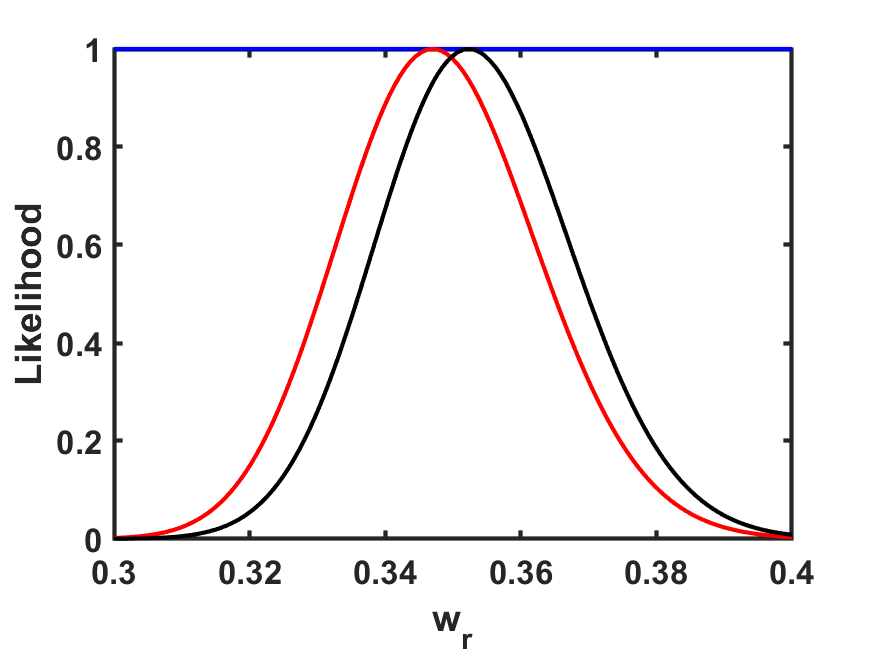}
    \caption{Constraints on the EMP model for flat universes with $\Omega_\Lambda=00$ and $w_r=const$; the top panels and bottom panels show the 2D and 1D (marginalized) constraints respectively. Blue, red and black lines correspond to one, two and three sigma constraints from cosmology, CMB and the joint data sets, and the colormap depicts the reduced chi.square.}
    \label{fig04}
  \end{center}
\end{figure*}

Figure \ref{fig03} and the fourth and final set of rows of Table \ref{table1} show the results of this analysis. Since $w_r$ only affects the temperature-redshift relation, the cosmological constraints on the matter density and $n$ are unchanged, but this data does not constrain $w_r$ at all. On the other hand, the CMB temperature data does not constrain the matter density, but it does constrain $n$ (with a sensitivity that is about a factor of two weaker than in the case with $w_r=1/3$) and also $w_r$ itself.

The combined constraints again force the model parameters to have values closer to the standard model ones, and despite this extra parameter the cosmology-only constraints are improved by about a factor of two for the matter density and a factor for three for the power of the nonlinear term. For the equation of state of radiation, there is no statistically significant deviation from the canonical value. We also note that the sensitivity of the $w_r$ constraint, of about 0.015, is comparable to that obtained recently in \cite{Gelo} for a different class of models.

\section{\label{conc}Conclusions}

We have provided improved low-redshift, background cosmology constraints on the general phenomenological class of EMP gravity models, in which the matter side of Einstein's equations includes a term proportional to some power, $n$, of the energy-momentum tensor. Taking notice that these models violate the temperature-redshift relation, we have relied on astrophysical measurements of the cosmic microwave background temperature, from the SZ effect and high-resolution optical spectroscopy of molecular and atomic species. 

Our analysis shows that this power is constrained to be about $|n|<0.01$ and $|n|<0.1$, depending on whether the canonical cosmological constant is assumed to vanish or is still allowed. Our analysis improves previous constraints on this parameter by more than a factor of three. Remarkably, the combination of CMB temperature measurements and cosmological data optimally breaks degeneracies between the power $n$ and the matter density, thereby improving constraints on the latter parameter by a factor of about two. Although we have restricted our analysis to low-redshift background cosmology data, both for the purpose of code validation and ease of comparison with previous works (and also of fair comparison between the two types of datasets), we note that even tighter constraints can presumably the obtained by the addition of high-reshhift data, e.g. from the CMB power spectrum.

It is clear that these EMP models are phenomenological models, useful as an illustration of how the recent acceleration of the universe could stem from the nonlinear term in a matter-only universe with $n=0$. Although this is technically not a cosmological constant, in practice it is observationally restricted to be indistinguishable from one. Note that this is true whether or not the canonical vacuum energy density $\Omega_\Lambda$, is present, though the constraints are much stronger (by about one order of magnitude) if $\Omega_\Lambda=0$.

That said, we must also emphasize that these models do not solve the so-called old cosmological constant problem of why it should be zero. Pragmatically, the main interest of these models is twofold. Firstly, they may be seen as illustrating a route for introducing a constant term in the Einstein equations---an effective cosmological constant. And secondly, they highlight the point that, whatever the physical mechanism underlying the recent acceleration of the universe, its large-scale gravitational behaviour is observationally constrained to be very close to that of a cosmological constant.

\begin{acknowledgments}
This work was financed by Portuguese funds through FCT (Funda\c c\~ao para a Ci\^encia e a Tecnologia) in the framework of the project 2022.04048.PTDC (Phi in the Sky, DOI 10.54499/2022.04048.PTDC). CJM also acknowledges FCT and POCH/FSE (EC) support through Investigador FCT Contract 2021.01214.CEECIND/CP1658/CT0001 (DOI 10.54499/2021.01214.CEECIND/CP1658/CT0001). 
\end{acknowledgments}
 
\bibliography{artigo}
\end{document}